\documentclass{Interspeech}

 \interspeechcameraready

\usepackage{multirow}
\title{FT-Boosted SV: Towards Noise Robust Speaker Verification for English Speaking Classroom Environments
}
\author[affiliation={1}]{Saba}{Tabatabaee}
\author[affiliation={2}]{Jing}{Liu}
\author[affiliation={1}]{Carol}{Espy-Wilson}

\affiliation{}{Department of Electrical and Computer Engineering, University of Maryland College Park}{USA}
\affiliation{}{Department of Education Policy, University of Maryland College Park}{USA}
\email{sabatb@umd.edu, jliu28@umd.edu, espy@umd.edu}

\usepackage{comment}

\begin{document}

\maketitle
\renewcommand{\thefootnote}{}%
\footnotetext{This work was supported by a Grand Challenge Award from the University of Maryland College Park.}%
\renewcommand{\thefootnote}{\arabic{footnote}} 
\begin{abstract}
Creating Speaker Verification (SV) systems for classroom settings that are robust to classroom noises such as babble noise is crucial for the development of AI tools that assist educational environments. In this work, we study the efficacy of finetuning with augmented children datasets to adapt the x-vector and ECAPA-TDNN to classroom environments. We demonstrate that finetuning with augmented children's datasets is powerful in that regard and reduces the Equal Error Rate (EER) of x-vector and ECAPA-TDNN models for both classroom datasets and children speech datasets. Notably, this method reduces EER of the ECAPA-TDNN model on average by half (a 5\% improvement) for classrooms in the MPT dataset compared to the ECAPA-TDNN baseline model. The x-vector model shows an 8\% average improvement for classrooms in the NCTE dataset compared to its baseline. 
\end{abstract}

\section{Introduction}

Developing speaker verification (SV) systems for classroom environments is crucial for fostering a more equitable learning atmosphere, while also providing teachers with a valuable tool to improve their teaching methods. Studies have primarily focused on developing SV systems for adult speech \cite{song2023dynamic,zhou2023adversarial,kim2024self,yu2020densely,jiang2019effective}. However, these systems may not be effective for handling children's speech due to the differences in speech properties between children and adults. Children's speech exhibits distinct acoustic and linguistic properties compared to adult speech \cite{Safavi2014ComparisonOS}. Moreover, it has been shown that children's speech exhibits more variability, both inter-speaker and intra-speaker, compared to adult speech \cite{Lee1999AcousticsOC}. The study by \cite{kaseva2021speaker} shows that training the SV system with both adult and children's datasets results in a 56\% improvement compared to training with the adult dataset alone, and a 45\% improvement compared to training with the children's dataset alone. 

Several research papers have investigated the effectiveness of finetuning pretrained models to develop SV systems for adults \cite{peng2023attention,sang2024efficient,peng2024fine} and children \cite{abed2024deep}. However, the effectiveness of finetuning method for developing SV systems in classroom environments, with their unique challenges, has not been thoroughly addressed. One challenge in classroom environments is babble noise, which involves the simultaneous speech of multiple children and complicates the verification process due to its similarity to the target speech. In this context, data augmentation with babble noise and other background noises can play a significant role in addressing this concern. Another challenge in developing SV systems for classrooms is the limited availability of children's datasets. In addition to finetuning pretrained models, which are models already trained on large datasets, with children's data, children's data augmentation can help address the issue of limited data availability.
 
Recent studies have introduced out-of-domain data augmentation methods to address the challenge of limited children's datasets in developing SV systems for children. In this approach, adult speech is modified by adjusting formants, pitch, and other acoustic properties to resemble children's speech \cite{singh2024childaugment, shahnawazuddin2020domain}. 

In the study by \cite{singh2024childaugment}, synthetic children's data was incorporated into the CSLU dataset to develop an SV system using the ECAPA-TDNN architecture. In another study \cite{shahnawazuddin2020domain}, they augmented the CMU kids dataset with synthetic data to train a SV system based on the x-vector model. However, children and adults produce speech differently, not only in terms of pitch and acoustic properties but also in pronunciation, grammar, and vocabulary complexity. Therefore, in this study, we focus on in-domain data augmentation to address the challenge of low-resource children's datasets, while also enhancing the robustness of SV systems to classroom-specific noise and conditions through two main steps: first, for development of classroom SV system, we combine three children's datasets that offer variability in age range and speech types (read, spontaneous, structured, and unstructured speech) that we can find in real classroom settings. We also conduct an ablation study to demonstrate the effectiveness of combining these three children datasets on the performance of SV systems in terms of Equal Error Rate (EER). Second, in addition to considering classroom background noises and reverberation effects found in public datasets such as MUSAN \cite{Snyder2015MUSANAM} and OpenSLR \footnote{https://www.openslr.org/28/}, we generate children's babble noise using existing children's datasets, as it is rarely available in public datasets. In this study, we propose that this augmentation approach is an effective way to enhance the robustness of SV systems to conditions and background noises typically found in the classroom domain. 

Previous studies have developed SV systems for Chinese-speaking classrooms \cite{Zheng2024NResNetNR}, and Indian-speaking schools with recorded audio in open environments \cite{Kadyan2022AutomaticSV}. However, to the best of our knowledge, this is the first study focused on SV systems for English-speaking classrooms, targeting real teaching sessions and student discussions, without role play, and accounting for diverse background noises, such as babble. In this paper, we present finetuning with augmented children's data as an effective approach to adapt x-vectors \cite{Snyder2018SpokenLR} and ECAPA-TDNN \cite{Desplanques2020ECAPATDNNEC} for SV in English-speaking classroom environments, enhancing their performance in real-world educational settings.

\textbf{Our key contributions} in this research are as follows:
\begin{itemize}
    \item We are the first to conduct research on SV systems for English-speaking classroom settings using real English-speaking classroom datasets.
    \item We perform an ablation study to evaluate the effectiveness of combining three children's datasets of MyST, CSLU, and CMU for developing classroom SV. 
    \item We conduct a comparative study of x-vector and ECAPA-TDNN models for classroom SV.
    \item We show that finetuning combined with our augmentation strategy is an effective tool to adapt x-vector and ECAPA-TDNN to noisy conditions like classrooms. As part of this augmentation method, we generate various children's babble noise, which closely resembles speech and is common in classroom environments. 
\end{itemize}
\section{Dataset Description}
For finetuning x-vector and ECAPA-TDNN models, we used three children's speech datasets: My Science Tutor (MyST), Center for Spoken Language Understanding (CSLU) kids speech, and Carnegie Mellon University (CMU) kids corpus, collectively referred to as MCC In this study. Each dataset was divided into 85\% for development (70\% training, 15\% validation) and 15\% for testing, with stratification based on speaker utterances to address data imbalance. After excluding recordings shorter than 3 seconds, the MCC development set included 161,722 number of utterances and 446.46 hours of speech from 2,544 speakers, and the testing set contained 28,540 number of utterances and 78.36 hours of speech from 2,540 speakers.

To evaluate the performance of the SV systems in the noisy classroom domain, we used English speaking classroom datasets from the National Center for Teacher Effectiveness (NCTE) and M-Powering Teachers (MPT) dataset. For NCTE and MPT, we employed Pyannote \cite{Bredin2023Pyannoteaudio2S, Plaquet2023PowersetMC} and TranscribeMe to identify speaker labels and determine the timestamps for the utterances as the ground truth. To ensure the SV systems maintain performance for adults during finetuning with children's datasets, we also used the Librispeech test-clean and Vox1-test adult datasets for evaluation. Detailed descriptions of each dataset are provided below.

\subsubsection{\textbf{MyST dataset}}
The MyST children's conversational speech corpus \cite{Pradhan2023MyST}, contains approximately 470 hours of English dialogue recorded from 1,371 students in grades 3 to 5 interacting with a virtual science tutor across eight science topics. After excluding audio recordings shorter than three seconds and those with errors during embedding extraction, the dataset was reduced to 1,354 children, 145,721 spoken utterances, and 435.87 hours of speech.

\subsubsection{\textbf{CSLU kids speech dataset}}

The CSLU dataset \cite{shobaki2007cslu} consists of spontaneous and prompted speech collected from 1,100 children ranging from Kindergarten to Grade 10. Each spontaneous speech session begins with alphabet recitation and includes a one-minute monologue. The dataset contains 39,804 spoken utterances and 80.17 hours of recorded speech.

\subsubsection{\textbf{CMU kids corpus dataset}}

The CMU dataset \cite{eskenazi1997cmu}, contains speech samples read aloud by children aged six to eleven. It includes 5,180 utterances from 76 speakers. After excluding recordings shorter than three seconds, the dataset consists of 4,737 utterances from 74 speakers, totaling 8.79 hours of speech.

\subsubsection{\textbf{Librispeech test-clean dataset}}

The Librispeech\cite{panayotov2015librispeech}, comprises adult English speech sourced from audiobooks. It includes a total of 5.4 hours of speech and 2,620 utterances from 40 speakers. In this work, we refer to the Librispeech test-clean dataset as LS.

\subsubsection{\textbf{Vox1-test dataset}}

The Vox1-test dataset is the test set of the VoxCeleb1 dataset \cite{nagrani2020voxceleb}, which contains interviews with celebrities uploaded to YouTube. This test dataset contains 11.2 hours of speech, with 4,874 utterances from 40 speakers. In this study, we refer to the Vox1-test dataset as Vox.

\subsubsection{\textbf{NCTE dataset}}

The NCTE dataset \cite{demszky2022ncte}, originally comprising 1,660 transcripts of 45-60 minutes of mathematics lessons for 4th and 5th grade elementary students.  Out of these recordings, 8 classrooms were randomly chosen for this study. After filtering out timestamps with no speech and excluding utterances shorter than 3 seconds, the final dataset includes 2.78 hours of speech, 2,918 utterances, and 97 speakers.

\subsubsection{\textbf{MPT dataset}}
The MPT is an in house dataset. We recorded six middle school mathematics classrooms. Similar to the NCTE dataset, these classrooms involved actual teaching and group discussions. At the beginning of each class, we conducted voice enrollment procedures in which both teachers and students read a short passage. We randomly selected two classroom recordings. After excluding recordings with no speech and utterances shorter than 3 seconds, the final dataset comprises 1.02 hours of speech, 465 utterances, and 30 speakers. Teachers wore a lanyard microphone. Swivl mics were used (one for the teacher and four scattered around the room to pick up student voices). The positioning of the microphones varied between classrooms, depending on their specific layouts and the audio streams were averaged.
\begin{table*}[ht!]
\caption{Comparison of SV models in terms of EER (\%) across three stages on test datasets: before finetuning, after finetuning with MCC data (FT), and after finetuning with  augmented MCC data (FT-Boosted). The lowest EER for each test set across all models is highlighted in bold.}
\centering 
\label{speaker_verification_models}
\begin{tabular}{|l|c|c|c|c|c|c|c|c|}
    \hline
    \centering \textbf{Model} & \multicolumn{8}{|c|}{\textbf{Test Dataset}} \\
    \cline{2-9} 
    & \textbf{\textit{MCC}}  & \textbf{\textit{MyST}} &  \textbf{\textit{CMU}} & \textbf{\textit{CSLU}}&\textbf{\textit{LS}} & \textbf{\textit{Vox}} & \textbf{\textit{Average of NCTE}}  & \textbf{\textit{Average of MPT}}\\
    \hline
    x-vector &  24.77 & 29.61 &14.50&23.04  & 8.61 &  8.40 & 21.97 & 15.74\\
    x-vector (FT) &  17.73 &21.86 &10.04&17.74  &  10.80 &  10.81  & 18.75 & 10.50\\
    x-vector (FT-Boosted) &  18.71 &23.61 & 10.18&17.20&  9.90  &  10.67 & 14.32 &  9.24\\
    \hline
    ECAPA-TDNN & 21.17 & 23.79&8.30 &19.85& \textbf{1.62} & \textbf{0.84} & 8.81 & 10.33\\
    ECAPA-TDNN (FT) & 16.03 & 18.75& 5.75& 14.46& 1.91  & 1.14 & 6.27&6.16\\
    ECAPA-TDNN (FT-Boosted) & \textbf{14.04} & \textbf{16.92} & \textbf{4.91}& \textbf{12.62}& 2.17  & 1.42  & \textbf{5.44}& \textbf{5.35} \\
    \hline
\end{tabular}
\end{table*}

\section{Methodology}
We finetuned two pretrained models of x-vector and ECAPA-TDNN that are available in the SpeechBrain toolkit \cite{ravanelli2021speechbrain} . These models were originally trained on the VoxCeleb1 and VoxCeleb2 datasets \cite{Chung2018VoxCeleb2DS}, which contain over 1,150,000 utterances from approximately 7,363 celebrities, including background noise, laughter, and overlapping speech.

All models have a sample frequency of 16 kHz. Therefore, we resampled audios to 16 kHz if the original sampling rate differed. The embedding dimensions for the x-vector and ECAPA-TDNN 
are 512 and 192
, respectively. We unfroze all model parameters (x-vector: 4,211,604; ECAPA-TDNN: 20,767,552
) and finetuned them using the MCC development set in two stages: first without an augmentation strategy, resulting in x-vector (FT) and ECAPA-TDNN (FT)
models; second with the augmented MCC data, resulting in x-vector (FT-Boosted) and ECAPA-TDNN (FT-Boosted)
models.

We used augmentation techniques to introduce variability characteristic of classroom environments into our training dataset, including background noise and reverberation. Background noises were selected from the MUSAN dataset and included construction sounds, the rustling of paper, knocking on a door, and door closing, etc. We simulated babble noise, typical in classroom settings, by combining audios from 12 to 25 randomly selected children from our development set. To make the models more robust to varying levels of noise in classroom environments, we randomly mixed background noises with the audio signals in the MCC training set at various signal-to-noise ratio (SNR) ranging from 5 dB to 15 dB. To further mimic classroom conditions, we simulated reverberation by convolving audio signals with various room impulse responses from the OpenSLR database.

During each epoch, the training dataset was shuffled, and the models were trained using the triplet loss function, which was popularized with the introduction of the FaceNet \cite{schroff2015facenet} and later used to develop SV systems \cite{zhang2017end,he2020text,zheng2023msranet}. The triplet loss aims to minimize the embedding distance between samples from the same speaker while maximizing the distance between samples from different speakers. In addition, our training strategy included online hard batch mining, which directs the network's attention to challenging samples that are difficult for the models to classify correctly. For training with triplet loss, we employed randomized selection of positive and negative pairs to promote better generalization across diverse data distributions.  We used the Adam optimizer and implemented early stopping with a patience of 8 epochs. Furthermore, we used a ‘ReduceLROnPlateau’ learning rate scheduler to reduce the learning rate with a patience of 8 epochs and decay of 0.5. The verification of speakers is performed by computing the cosine distance between speaker embeddings, and after applying score normalization, the EER is calculated.

\section{Results and Discussion}
\begin{table*}[t!]
\caption{Detailed results of SV systems in EER (\%) for each classroom of NCTE and MPT dataset. The lowest EER for each test set across all models is highlighted in bold.}
\centering 
\label{speaker_verification_models}
\begin{tabular}{|l|c|c|c|c|c|c|c|c|c|c|}
    \hline
        \centering \textbf{Model} & \multicolumn{8}{|c|}{\textbf{NCTE}} & \multicolumn{2}{|c|}{\textbf{MPT}}\\
    \cline{2-11} 
    &\textbf{\textit{31}}& \textbf{\textit{201}} & \textbf{\textit{230}} & \textbf{\textit{2757}} & \textbf{\textit{4106}}& \textbf{\textit{4191}}& \textbf{\textit{4352}}& \textbf{\textit{4651}} & \textbf{\textit{ES}}& \textbf{\textit{MN}}\\
    \hline
    
    x-vector & 31.82 & 12.05 & 16.82 & 30.77 & 27.03 & 26.92 & 12.46 & 17.90 & 11.42 & 20.05\\ 
    x-vector (FT) & 36.91 & 5.88 & 10.52 & 25.00 & 24.27 & 26.28 & 7.48 & 13.62 & 6.85 & 14.15\\ 
    x-vector (FT-Boosted) & 31.70 & 5.11 & 7.08 & 17.67 & 18.99 & 19.46 & 5.96 & 8.57 & 6.05 & 12.43\\ 
    \hline
    ECAPA-TDNN & 13.86 & 5.86  & 4.25& 11.75 & 16.04 & 8.23 & 3.43 & 7.07 & 7.81 & 12.85\\ 
    ECAPA-TDNN (FT) & 9.26 & 4.39 & 2.88 & 8.09 & 12.22 & 7.05 & 2.05 & 4.18 & 4.33 & 7.99\\ 
    ECAPA-TDNN (FT-Boosted) & \textbf{8.52} & \textbf{3.83} & \textbf{2.86} & \textbf{7.67} & \textbf{9.64} & \textbf{6.53} & \textbf{1.15} & \textbf{3.33} & \textbf{3.58} & \textbf{7.12}\\ 

    \hline
\end{tabular}
\end{table*}

\subsection{Analysis of finetuning x-vector and ECAPA-TDNN}
Table 1 shows the EER across all test folds for the x-vector and ECAPA-TDNN models. To evaluate the SV systems on children's datasets, we use the MCC, which is the combination of test sets from the MyST, CSLU, and CMU datasets. Additionally, we report the EER for each individual test set  of MyST, CSLU, and CMU separately. For the off-the-shelf x-vector and ECAPA-TDNN pretrained models, we observe high EERs for the children’s test datasets (MCC, MyST, CSLU, and CMU) as well as the classroom datasets (NCTE and MPT). These high EER values can be attributed to the fact that the models were initially trained on the VoxCeleb1 and VoxCeleb2 datasets, which mainly consist of adult celebrity interviews. Consequently, these pretrained models struggle to effectively handle the distinct characteristics of children’s speech, as well as the noise and variable conditions typical of classroom environments.

Finetuning the x-vector and ECAPA-TDNN pretrained models with the MCC dataset (referred to as x-vector (FT) and ECAPA-TDNN (FT)) improves the performance of both models on the children’s test datasets and classroom datasets. This result emphasizes the importance of domain-specific finetuning in adapting models to the unique acoustic features of children's speech. It suggests that even before introducing background noise to the MCC dataset, training on domain-relevant data enhances model robustness and performance, making them more suitable for SV tasks in environments with children.

The results for x-vector (FT-Boosted) and ECAPA-TDNN (FT-Boosted) demonstrate the effectiveness of finetuning the pretrained x-vector and ECAPA-TDNN models with the augmented MCC development set, boosted with classroom background noises such as babble noise. By incorporating these environmental factors during training, we achieved significant performance improvements on classroom test sets. The x-vector (FT-Boosted) model showed a notable reduction in EER for both the NCTE (a 7.65\% reduction in EER) and MPT (a 6.5\% reduction in EER). The ECAPA-TDNN (FT-Boosted) model showed significant relative improvements of 48\% for the MPT classroom dataset (approximately cutting the EER by half) and 38\% for the NCTE dataset. These results suggests that domain-specific finetuning with relevant background noise improves the robustness of SV systems in noisy environments like classrooms.

During the finetuning process, we closely monitored the performance of the models on adult speech (using the Vox and LS test sets). This was particularly important, as the teacher-to-student speech ratio in classroom environments can vary, and maintaining good performance for adult speech (teachers) is crucial. Based on the results presented in Table 1, and Table 2, the ECAPA-TDNN (FT-Boosted) achieved the top performance in terms of EER for both children's speech and classroom. Furthermore, this model retains its strong performance on adult speech, demonstrating its robustness in handling speech from both adults and children, as well as in noisy classroom conditions.
\subsection{Detailed analysis of SV in classrooms}
In this section, we provide a detailed analysis of the results for classroom environments, as presented in Table 2. Looking at the results, it is clear that finetuning with the augmented MCC dataset improves performance across all classrooms. The variation of EER in Table 2, both high and low, across different classroom recordings can be attributed to the unique characteristics of each recording, leading to performance differences. Although we consider classroom recordings as a single domain, the specific type of classroom setting—whether collaborative or instructional—can influence factors such as noise levels and the teacher-to-student speech ratio, all of which impact the EER. Another important factor is the microphone configuration, which can impact SV system performance. For instance, recording 31 in NCTE dataset is the only one captured using a far-field microphone. In this challenging scenario, the x-vector (FT-Boosted) model showed only a slight improvement, whereas the ECAPA-TDNN (FT-Boosted) model achieved a 5.34\% reduction in EER compared to its baseline, and a 23.18\% improvement over the x-vector (FT-Boosted) model. This highlights the greater generalizability of ECAPA-TDNN (FT-Boosted) to different microphone configurations. 

One key factor observed across all classrooms is the presence of babble noise, which varies depending on the classroom setting. In some classrooms (2757, 4106, 4191, and 4651 in NCTE dataset, and MN recording in MPT dataset) the teacher assigns students to work in groups and engage in discussions, leading to higher levels of babble noise. This created a much noisier environment compared to other classrooms, with an elevated degree of children's babble noise. Despite this challenge, the ECAPA-TDNN (FT-Boosted) model demonstrated a significant reduction in EER. For instance, in recording 4106, the ECAPA-TDNN (FT-Boosted) model reduced the EER by 6.40\% compared to its baseline and by 9.35\% compared to the x-vector (FT-Boosted) model. This suggests the enhanced capability of ECAPA-TDNN (FT-Boosted) to handle babble noise in classroom environments.

\subsection{Analysis of the choice of development set}
The results of our ablation study, presented in Table 3, highlight the significant impact of dataset choice on the performance of the ECAPA-TDNN (FT-Boosted) model, which emerges as the top-performing SV system for both children's speech and classroom environments (as shown in Tables 1 and 2). We began our experiments by finetuning the model with the MyST dataset alone, since it is considerably larger than both the CSLU and CMU datasets. In each subsequent step, we finetuned the ECAPA-TDNN (FT-Boosted) model with different combinations of datasets, evaluating its performance on both the MPT and NCTE classroom datasets. As shown in Table 3, adding either the CMU or CSLU dataset to the MyST dataset during finetuning leads to a reduction in the EER for both the NCTE and MPT datasets, compared to using MyST alone. Finally, the results demonstrate that finetuning with the combination of MyST, CSLU, and CMU datasets (referred to as MCC) achieves the lowest EER for both the NCTE and MPT datasets.

The MCC dataset showed to be a highly effective choice for SV in classroom environments due to its diverse range of speech styles, including read, conversational, structured, and unstructured speech, as well as its broad age coverage. This variety enables the model to better handle the variations in speech patterns and age, enhancing its robustness and performance across different classroom scenarios. As demonstrated by the results in Table 3, finetuning with the MCC dataset significantly improves the model’s ability to accurately verify speakers in real-world classroom settings, making it an optimal dataset for this task.

\begin{table}[h!]

\caption{Average EER (\%) for the ECAPA-TDNN (FT-Boosted) model across all classrooms in the NCTE and MPT datasets. The lowest EER for each dataset across all development sets is highlighted in bold.}
\centering 
\label{speaker_verification_models}
\begin{tabular}{|l|c|c|}
\hline
\multirow{2}{*}{Finetuning Dataset} & \multicolumn{2}{c|}{ECAPA-TDNN (FT-Boosted)} \\ \cline{2-3}
 & \textit{Average of NCTE} & \textit{Average of MPT} \\ \hline
MyST & 6.24 & 5.79 \\ 
MyST + CMU & 6.09 & 5.66 \\ 
MyST + CSLU & 5.63 & 5.52 \\
MCC & \textbf{5.44} & \textbf{5.35} \\  \hline
\end{tabular}
\end{table}

\section{Conclusions And Future Work}
In this study, we explored speaker verification in real English-speaking classrooms, where both children and adults are present. We demonstrated the effectiveness of finetuning pretrained models using a domain-specific, augmented dataset for speaker verification tasks in noisy and challenging environments, such as classrooms. Our results show that this approach significantly improves the performance of both the ECAPA-TDNN and x-vector models for children's speech and classroom environments. We also found that the ECAPA-TDNN (FT-Boosted) model exhibited superior robustness, especially in handling variations in microphone configurations and classroom noise. Additionally, we propose that the optimal development set for finetuning speaker verification models is a combination of the MyST, CSLU, and CMU datasets, due to their diverse age range and speech styles. This diverse combination enables speaker verification models to generalize more effectively, improving their performance in real-world educational settings. 

Future work will involve conducting larger finetuning trials with additional classroom recordings, as well as using self-supervised feature extraction techniques. This approach aims to further enhance the robustness and generalizability of speaker verification systems in classroom environments.

\bibliographystyle{IEEEtran}
\bibliography{mybib}

\end{document}